\documentclass[12pt]{article}
\usepackage{fullpage}
\usepackage[utf8]{inputenc}
\usepackage{amsmath, biometri, mathtools}
\usepackage{amssymb}
\usepackage{amsfonts,}
\usepackage{bbm}
\usepackage{graphicx}
\usepackage{float}
\usepackage{caption}
\usepackage{subcaption,mathrsfs,multirow}
\usepackage[ruled]{algorithm2e}
\usepackage{array}

\usepackage{xr}
\usepackage{verbatim}
\usepackage{listings}
\usepackage{color}
\usepackage{epsfig}
\usepackage{wasysym}

\newcommand{\bbeta}{\boldsymbol{\beta}}
\newcommand{\Beta}{\boldsymbol{\eta}}

\newcommand{\bxi}{\boldsymbol{\xi}}

\usepackage{color}

\newcommand{\comout}[1]{}

\def\Asc{\mathcal{A}}

\def\lpbox#1{\vskip1mm \begin{center}
        \hspace{.0\textwidth}\vbox{\hrule\hbox{\vrule\kern6pt
\parbox{.95\textwidth}{\kern6pt \blue #1 (LP)\vskip6pt}\kern6pt\vrule}\hrule}
        \end{center} \vskip-5mm}

\usepackage[usenames,dvipsnames,svgnames,table]{xcolor}
\usepackage[colorlinks=true,
            linkcolor=red,
            urlcolor=blue,
            citecolor=blue]{hyperref}

\graphicspath{{./}{./Figures/}}

\usepackage{setspace,tcolorbox}
\definecolor{lbcolor}{rgb}{0.95,0.95,0.95}
\def\ESE{\mbox{ESE}}
\def\ASE{\mbox{ASE}}

\def\bZ{\mathbf{Z}}
\def\bz{\mathbf{z}}

\def\Z{\mbox{Z}}
\def\bG{\mathbf{G}}

\definecolor{darkred}{RGB}{150,50,50}
\definecolor{brown}{RGB}{250,100,100}
\definecolor{green}{RGB}{000,150,100}
\definecolor{purple}{RGB}{200,000,250}

\def\blue{\color{blue}}

\def\blue{\color{blue}}

\def\trans{^{\scriptscriptstyle \sf T}}

\begin{document}

\begin{center}
{\large \bf Semiparametric Joint Modeling to Estimate the Treatment Effect on a Longitudinal Surrogate with Application to Chronic Kidney Disease Trials} \vspace{.1in}

Xuan Wang \\
{\em Division of Biostatistics, Department of Population Health Sciences, University of Utah, SLC, UT 84108, USA} \\ xuan.wang@utah.edu \\[1ex]

Jie Zhou \\
{\em School of Mathematics, Capital Normal University, Beijing 100048, China}\\[1ex]

Layla Parast \\
{\em Department of Statistics and Data Science, University of Texas at Austin, Austin, TX 78712, USA} \\[1ex]

Tom Greene \\
{\em Division of Biostatistics, Department of Population Health Sciences, University of Utah, SLC, UT 84108, USA} \\[1ex]

%
%

\end{center}

\begin{abstract}
In clinical trials where long follow-up is required to measure the primary outcome of interest, there is substantial interest in using an accepted surrogate outcome that can be measured earlier in time or with less cost to estimate a treatment effect. For example, in clinical trials of chronic kidney disease (CKD), the effect of a treatment is often demonstrated on a surrogate outcome, the longitudinal trajectory of glomerular filtration rate (GFR). However, estimating the effect of a treatment on the GFR trajectory is complicated by the fact that GFR measurement can be terminated by the occurrence of a terminal event, such as death or kidney failure. Thus, to estimate this effect, one must consider both the longitudinal outcome of GFR, and the terminal event process. Available estimation methods either impose restrictive parametric assumptions with corresponding maximum likelihood estimation that is computation-intensive or other assumptions not appropriate for the GFR setting. In this paper, we build a semiparametric framework to jointly model the longitudinal outcome and the terminal event, where the model for the longitudinal outcome is semiparametric, and the relationship between the longitudinal outcome and the terminal event is nonparametric. The proposed semiparametric joint model is flexible and can be extended to include nonlinear trajectory of the longitudinal outcome easily. An estimating equation based method is proposed to estimate the treatment effect on the slope of the longitudinal outcome (e.g., GFR slope). Theoretical properties of the proposed estimators are derived. Finite sample performance of the proposed method is evaluated through simulation studies. We illustrate the proposed method using data from the Reduction of Endpoints in NIDDM with the Angiotensin II Antagonist Losartan (RENAAL) trail to examine the effect of Losartan on GFR slope.

\vspace{10pt}

\noindent {\it{Keywords}}: clinical trial; longitudinal outcome; surrogate, terminal event; joint modeling; shared parameter model; semiparametric model 
\end{abstract}

\setstretch{1.5}

\newpage
\section{Introduction}

Randomized clinical trials, particularly trials, of chronic kidney disease (CKD), often require long-term follow up of participants in order to evaluate the effectiveness of novel treatments or interventions. As pressure from patients, caregivers and policymakers to expedite decision making of new treatments has grown, there has been substantial interest in using surrogate biomarkers that can be measured earlier in time or with less cost to make decisions about the treatment effect. For example, in clinical trials of CKD, the longitudinal trajectory of glomerular filtration rate (GFR) has been shown to be associated with kidney disease and is considered by the US Food and Drug Administration to be an accepted surrogate endpoint for traditional drug approval (\cite*{FDA}). That is, the efficacy of a new drug/treatment can be demonstrated by showing that there is a treatment effect on the GFR trajectory, which can potentially be observed earlier than a long-term clinical event such as the composite of kidney failure, sustained GFR$<$15 ml min$^{-1}$ per 1.73 m$^2$, or doubling of serum creatinine.

Unsurprisingly, estimating the effect of a treatment on the GFR trajectory can be difficult because GFR measurement can be (and often is) terminated by the occurrence of a terminal event, such as death or kidney failure. Thus, to estimate this effect, one must consider both the longitudinal outcome of GFR, and the event process. Often, such estimation is based on shared parameter models (\cite*{vonesh2006shared,vonesh2019mixed,inker2023meta}). The shared parameter models for GFR trajectory involve jointly modeling trends (ie, slopes) in GFR over time together with potentially informative termination by the terminal event. This consists of jointly fitting a mixed-effect model for GFR and an event time model for the terminal event using a set of shared random effects. This approach can handle the challenges such as different early and late treatment effects on the slope, within-subject and between-subject heteroscedasticity, and informative termination resulting from patient dropout due to death or kidney failure. However, estimation within this approach is often based on a parametric likelihood, which imposes several model assumptions with many unknown parameters. Specifically, the shared parameter models involve modeling the longitudinal outcome parametrically, and the relationship between the longitudinal outcome and the terminal event parametrically. Such model assumptions are difficult to verify and model misspecification would likely result in bias. In addition, computation for these models is usually intensive, and achieving model convergence can be challenging.

Certainly, there exist some useful semiparametric methods for the analysis of longitudinal data that do not require strict parametric assumptions and are easier to implement (\cite*{lin2001semiparametric,sun2005semiparametric,sun2007regression}). While these methods require less parametric assumptions, they generally have other limitations that make them unsuitable for the GFR setting. For example, \cite{jin2006analysis} proposed an approach to model the longitudinal outcome and the survival time, separately, assuming they were independent conditional on baseline covariates. However, in many applications, including the analysis of the GFR trajectory, the longitudinal outcome may still be related to the event, even after adjusting for other covariates. As another example, \cite{sun2012joint} proposed to model the longitudinal outcome in the presence of a dependent terminal event via latent variables.  Unfortunately, the model involved conditioning on both the latent variables and the survival time being greater than the time of interest, making the practical interpretation of the resulting parameters unclear. 

In this paper, motivated by the desire to estimate the effect of a treatment on GFR trajectory, we propose a semiparametric joint modeling framework for a longitudinal outcome and a terminal event that overcomes the limitations of existing methods. Our modeling framework is flexible as it can handle heteroscedasticity, informative termination from terminal events, and it includes the parametric fixed-effect models of \cite{vonesh2006shared} and \cite{vonesh2019mixed} as special cases. Importantly, it fully captures the relationship between the longitudinal outcome and the terminal event, while allowing the association between the longitudinal outcome and the terminal event to be unspecified or nonparametric. An easy-to-implement estimating equation approach is developed for parameter estimation, including for estimation of the treatment effect on the longitudinal outcome slope, the ultimate quantity of interest. 
The resulting estimators are shown to be consistent and asymptotic normal and simulation studies are conducted to demonstrate the performance of the proposed methods. In addition, we apply our approach to analyze data from the Reduction of Endpoints in NIDDM with the Angiotensin II Antagonist Losartan (RENAAL) trial to examine the effect of Losartan on GFR slope.

\section{Semiparametric Joint Modeling (SJM) for the Longitudinal Outcome and the Terminal Event   \label{optimal}}

\subsection{Notation and Model \label{notation}}

Let $Y(t)$ denote the longitudinal outcome at time t. Let $\bZ$ be the $p \times 1$ vector of baseline covariates, including the binary treatment indicator $A$ as its first component, $\bZ=(A,\bZ_{-1}^{\trans})^{\trans}$ and $\bZ_{-1}$ is simply $\bZ$ without the first component $A$. In addition, let $D$ be the time of the terminal event and $C$ be the censoring time. Let $T = C\wedge D$ and $\Delta = I(D < C)$, where $C \perp D | \bZ$, $a \wedge b = min(a, b)$ and $I(\cdot)$ is the indicator function. Let $N(t)$ be the counting process denoting the number of the observation times before or at time t. For a random sample of $n$ subjects, the observed data consist of $\{Y_i(t)dN_i(t), T_i, \delta_i, \bZ_i, N_i(t), 0 < t < T_i, i=1, ..., n\}$. In this section, we propose a semiparametric joint modeling (SJM) approach that is flexible and robust, though still assumes a linear model for $Y(t)$; in Section \ref{expand}, we relax this assumption via nonlinear regression. 

We first assume a Cox proportional hazards model for the hazard function of the event time $D$:
\begin{eqnarray}
\lambda_0(t|\bZ) = \lambda_0(t)\exp(\Beta^{\trans} \bZ), \label{cox}
\end{eqnarray}
where $\lambda_0(t)$ is an unspecified baseline hazard function, $\Beta$ is a vector of unknown regression parameters with true value $\Beta_0$, we will use $\eta_{A}$ to denote the coefficient associated with treatment $A$ i.e., quantifying the treatment effect on the terminal event. Model (\ref{cox}) is the marginal model for the terminal event $D$; below, we introduce the model for the longitudinal marker $Y(t)$, which will include the relationship between $D$ and $Y(t)$. 

For the observation process, it is assumed that $N(t)$ is a non-stationary poisson process that follows the marginal model:
\begin{eqnarray}
E\{dN(t)\}= d\mu_0(t), \label{poisson}
\end{eqnarray}
where $\mu_0(t)$ is the intensity function or rate function with $\mu_0(0)=0$. 
This formulation for the visit process is quite general, and accommodates prospective studies with fixed visit schedules by stipulating that $\mu_0(t)$ is non-zero only during time intervals when visits occur, such as the kidney clinical trials where the patients had regularly scheduled visit times. Model (\ref{poisson}) will be used to handle the nonparametric component in the model for the longitudinal marker $Y(t)$ introduced below.

Next, we assume that the longitudinal process $Y(t)$ is related to the covariates by the following flexible model:
\begin{eqnarray}
Y(t)&=& \alpha_0(t,v)+\beta_{0}A+\beta_{1}A t+\bbeta_2^{\trans} \bZ_{-1}+e(t) \nonumber\\
&=&\alpha_0(t,v)+ {\bbeta}^{\trans}\tilde{\bZ}(t) +e(t), \label{gfr}
\end{eqnarray}
where ${\bbeta}=((\beta_0,\beta_1)^{\trans},\bbeta_2^{\trans})^{\trans}$, $\tilde{\bZ}(t)=(A, A t, \bZ_{-1}^{\trans})^{\trans}$, $v$ is a latent variable that is independent of $\bZ$ but may be associated with $D$ {through $\varepsilon=\log \Lambda_0(D)+\Beta_0^{\trans}\bZ$ which follows the extreme value distribution}, $\alpha_0(t,v)$ is an unknown function, and $e(t)$ is a mean zero process independent of other variables. The covariate vector $\bZ_{-1}$  may include covariates related to the terminal event and/or the longitudinal process. The function $\alpha_0\left(t,v\right)$ is an unspecified function of $t$ which contains, for example, an overall intercept and slope for GFR in the control group (such that $\beta_0$ and $\beta_1$ represent treatment effects on the intercept and slope, respectively). Furthermore, $\alpha_0(t,v)$ will typically include random effects (such as random intercepts and slopes) that characterize variation in GFR trajectories between subjects, and these random effects can be related to D through $\varepsilon$. The mean-zero process $e(t)$ may have constant variance over time or varying variance (heteroscedasticity) over time, and $e(t)$ at different time points may be correlated. This joint model is semiparametric and very flexible, and it includes the widely used parametric model as special cases (\cite*{vonesh2006shared,vonesh2019mixed,inker2023meta}). Importantly, $\beta_{1}$ is the treatment effect on the slope of the longitudinal outcome which is our ultimate quantity of interest.

\noindent {\bf Remark 1.} Patterns of change in GFR following initiation of an intervention are often nonlinear, with possibly different directions and rates of change in early follow-up (acute slope) versus longer-term follow-up (chronic slope). Model (\ref{gfr}) can easily be modified to consider both the acute treatment effect, $\beta_{1}$, and the chronic treatment effect, $\beta_{1}+\beta_{2}$, on the  surrogate biomarker slope, as follows. 
\begin{eqnarray}
Y(t)&=& \alpha_0(t,v)+\beta_{0}A+\beta_{1}A t+\beta_{2}A ((t-t^*)\vee 0) +\bbeta_3^{\trans} \bZ_{-1}+e(t) \nonumber\\
&=&\alpha_0(t,v)+ {\bbeta}^{\trans}\tilde{\bZ}(t) +e(t), 
\end{eqnarray}
where $t^*$ is a change point that the slope may be different before and after this time point, ${\bbeta}=((\beta_0,\beta_1,\beta_2)^{\trans},\bbeta_3^{\trans})^{\trans}$, $\tilde{\bZ}(t)=(A, A t, A((t-t^*)\vee 0), \bZ_{-1}^{\trans})^{\trans}.$

\noindent {\bf Remark 2.} The longitudinal marker $Y(t)$ is a linear function of the covariate vector $\bZ$ conditional on $\bZ$ and $v$. $Y(t)$ may be related to the terminal event $D$ through $v$ even conditional on the covariates $\bZ$, which induces the informative termination. This informative termination cannot be handled by \cite{jin2006analysis}, which assumes $Y(t)$ is independent of $D$ conditional on the covariates $\bZ$. Model (\ref{gfr}) is the model for the longitudinal outcome, $Y(t)$, where the coefficients of $A$ and $At$ have causal interpretation. Since the model of \cite{sun2012joint} is for $Y(t)$ conditional on the subject still under observation (not terminated), their coefficients no longer have a causal interpretation. 


\subsection{Estimation \label{estimation}}
We first estimate the unknown parameters in the Cox model (\ref{cox}) based on existing methods. Next, to estimate the unknown parameter ${\bbeta}$ in model (\ref{gfr}), we propose to borrow strength from model (\ref{poisson}) for the observation process and estimate ${\bbeta}$ based on an estimating equation, constructed from a mean zero process. We will show that the estimating equation has a closed form solution. Note that we do not need to estimate the unknown function $\alpha_0(t,v)$ in model (\ref{gfr}) so the form of $\alpha_0(t,v)$ can be arbitrary/nonparametric. Below, we detail the estimation method.

Let $\widehat{\Beta}$ be the maximum partial likelihood estimator of $\Beta$ and $\widehat{\Lambda}_0$ be the Breslow estimator of the baseline cumulative hazard function $\Lambda_0$. By direct derivations, shown in the Appendix, we have
\begin{eqnarray}
&&E[ \{Y(t)-{\bbeta}^{\trans}\tilde{\bZ}(t)\}dN(t)|\bZ,T>t]= dH(t,\log \Lambda_0(t)+\Beta^{\trans}\bZ), \nonumber\\
\text{where}\ && dH(t,s)={E\left[ d\Asc(t,v) |\varepsilon>s  \right] }\nonumber\\
&&=  \frac{E[\{Y(t)-{\bbeta}^{\trans}\tilde{\bZ}(t)\}dN(t) I(\log \Lambda_0(T)+\Beta^{\trans}\bZ>s>\log \Lambda_0(t)+\Beta^{\trans}\bZ)]}{E[ I(\log \Lambda_0(T)+\Beta^{\trans}\bZ>s>\log \Lambda_0(t)+\Beta^{\trans}\bZ)]},\label{dH}
\end{eqnarray}
where $d\Asc(t,v)=\alpha_0(t,v)d\mu_0(t)$. As a result, the process  $$dM_i(t, {\bbeta}):=I(T_i>t)\left[\{Y_i(t)-{\bbeta}^{\trans}\tilde{\bZ}_i(t)\}dN_i(t)- \{d \bar{Y}_i(t)-{\bbeta}^{\trans}d \bar{{\bG}}_i(t)\} \right]$$ is approximately zero for given covariates, and 
$$d\bar{Y}_i(t)=\frac{\sum_{j=1}^n \phi_j(t,\bZ_i) Y_j(t)dN_j(t) }{\sum_{j=1}^n \phi_j(t,\bZ_i)}, d\bar{\bG}_i(t)=\frac{\sum_{j=1}^n \phi_j(t,\bZ_i) \tilde{\bZ}_j(t)dN_j(t) }{\sum_{j=1}^n \phi_j(t,\bZ_i)},$$
$$\text{and}\ \phi_j(t,\bZ_{i})=I\left(\log \widehat \Lambda_0(T_j)+ \widehat\Beta^{\trans}\bZ_{j}>\log \widehat \Lambda_0(t)+ \widehat\Beta^{\trans}\bZ_{i}>\log  \widehat \Lambda_0(t)+ \widehat\Beta^{\trans}\bZ_{j}\right).$$
Based on this, we construct an estimating equation for ${\bbeta}$ as:
\begin{eqnarray}
U({\bbeta})\!=\!\sum_{i=1}^n \!\int_0^{\tau}\!\!\! I(T_i>t)\left[\tilde{\bZ}_i(t)\!-\!\bar{{\bZ}}_i(t)\right]\left[Y_i(t)dN_i(t)\!-\!d\bar{Y}_i(t)\!-\!\{{\bbeta}^{\trans}\tilde{\bZ}_i(t)dN_i(t)\!-\!{\bbeta}^{\trans}d\bar{\bG}_i(t) \}\right]\!\!, \label{estimating}
\end{eqnarray}
{where $\tau$ denote the end of the study, and } $$\bar{{\bZ}}_i(t)=\frac{\sum_{j=1}^n \phi_j(t,\bZ_i)\tilde{\bZ}_j(t) }{\sum_{j=1}^n \phi_j(t,\bZ_i)}.$$
Solving $U({\bbeta})=0$ we obtain closed form solution:
\begin{eqnarray}
\widehat{\bbeta}&=&\left[\sum_{i=1}^n \int_0^{\tau}  [\tilde{\bZ}_i(t)-\bar{{\bZ}}_i(t)] I(T_i>t) \{\tilde{\bZ}_i(t)dN_i(t)-\bar{\bG}_i(t)\}\right]^{-1}
 \nonumber\\
&& \times\left[ \sum_{i=1}^n \int_0^{\tau} [\tilde{\bZ}_i(t)-\bar{{\bZ}}_i(t)] I(T_i>t) \{Y_i(t)dN_i(t)-\bar{Y}_i(t)\} \right].
\end{eqnarray}

\noindent {\bf Remark 3.} The covariance of $Y(t)$ may change as time increases; in this case, we may construct a weighted estimating equation by adding weight to the estimating equation (\ref{estimating}) to get a more efficient estimator of $\bbeta$. The weight maybe related to the inverse of the covariance of $Y(t)$.

\noindent {\bf Remark 4.} Our model and inference method can be extended to incorporate external time-dependent covariates $\mathbf{Z}(t)$ in the above formulation. In particular, when $\mathbf{Z}(t)$ is time-dependent, the model (\ref{cox}) for the terminal event becomes $\Lambda\left(t |\mathbf{Z}\right)=G\left[\int_{0}^{t}\exp\{-\Beta^{\trans} \bZ(s)\}d\Lambda(s)\right]$, where $\Lambda\left(t |\mathbf{Z}\right)$ is the conditional hazard function of T given $\bZ$, and $G(x)$ is a given transformation function. The weight in $\bar{{\bZ}}_i(t), \bar{Y}_i(t), \bar{\bG}_i(t)$ can be modified accordingly.

\subsection{Asymptotic Properties \label{asymptotics}}
 \noindent{\bf Theorem 1.} \ {\it Under the regularity conditions (C1)-(C3)
  stated in the Appendix,
  $n^{1/2} ( \widehat \bbeta - \bbeta_0 )$
  has asymptotically a  normal distribution with mean zero and
  covariance matrix $A^{-1}\Sigma (A^{\trans})^{-1}$,
  where $\bbeta_0$ is the true value of $\bbeta$, $A$ and $\Sigma$ are
  defined in the Appendix.
 }
 
We outline the proof of Theorem 1 in the Appendix. Since the closed form of $\Sigma$ is difficult to estimate in practice, we estimate the asymptotic variance using perturbation resampling in our numerical studies (\cite*{parast2016robust,wang2020model}). 

\section{Semiparametric Joint Modeling with Nonlinear Regression for the Longitudinal Outcome} \label{expand}
While our SJM approach in Section \ref{optimal} is flexible and robust, it still assumes that the GRF trajectory is linear or piecewise linear in model (\ref{gfr}). The linearity assumption for the GRF trajectory, while common in the literature, is likely to be incorrect (\cite*{li2012longitudinal,xie2016estimated}). Fortunately, our approach can be adapted to accommodate a nonlinear relationship easily, while such an adaptation is hard or infeasible under the traditional parametric framework.

Specifically, for the event time, we assume the same Cox PH model as model (\ref{cox}), while for the longitudinal outcome, $Y(t)$, we assume that $Y(t)$ follows 
\begin{eqnarray}
Y(t)&=& \alpha_0(t,v)+\beta_{0}A+g(t)A +\bbeta_2^{\trans} \bZ_{-1}+e(t), \label{gfr2}
\end{eqnarray}
where $g(t)$ is an unknown function of $t$, which can be approximated by a linear combination of spline bases of the time t, $\bbeta_{1}^{\trans} \phi(t)$. Thus, 
\begin{eqnarray}
Y(t)&\approx& \alpha_0(t,v)+\beta_{0}A+\bbeta_{1}^{\trans} \phi(t) A +\bbeta_2^{\trans} \bZ_{-1}+e(t) \nonumber\\
&:=&\alpha_0(t,v)+ {\bbeta}^{\trans}\tilde{\bZ}(t) +e(t), \label{gfr3}
\end{eqnarray}
where ${\bbeta}=(\beta_0,\bbeta_1^{\trans},\bbeta_2^{\trans})^{\trans}$, $\tilde{\bZ}(t)=(A, A \phi(t)^{\trans}, \bZ_{-1}^{\trans})^{\trans}$. That is, model (\ref{gfr3}) is an extension of model (\ref{gfr}) where we use a spline basis expansion of the time $t$, $\phi(t)$, to allow the longitudinal outcome to vary nonlinearly across time. 

Since the association between $Y(t)$ and treatment is more complex here because we allow the treatment effect to be nonlinear in time, it is not as straightforward to simply point out a regression coefficient that equals the treatment effect on $Y(t)$. Instead, it is helpful to consider the formal definition of the treatment effect on $Y(t)$ using potential outcomes notation to identify the quantity of interest: $$\Delta_{Y}(t)=E[Y^{(1)}(t)-Y^{(0)}(t)]=\beta_{0}+g(t),$$ where $Y^{(a)}(t), a=1, 0$ is the potential outcome at $t$ if the individual received treatment $A=a$, and may also be nonlinear in time. The treatment effect on the slope of the longitudinal outcome, instead of $\beta_1$ in model (\ref{gfr}), is $$\Delta_{Yslope}(t)=E[Y^{(1)}(t)-Y^{(1)}(0)-\{Y^{(0)}(t)-Y^{(0)}(0)\}]/t=g(t)/t. $$

For estimation of $g(t)$, we can use a similar estimating equation based method as in Section \ref{optimal} by replacing the covariates in time $t$ with the spline basis of the time $t$, denoting the resulting estimator as $\widehat{g}(t)=\widehat{\bbeta}_{1}^{\trans} \phi(t)$. The asymptotic properties of $\widehat{\bbeta}_{1}$ can be established similar to Section \ref{asymptotics}. The nonparametric estimator $\widehat{g}(t)$ is consistent for the true function $g(t)$ but the convergence rate of $\widehat{g}(t)$ is slower than the usual convergence rate for parametric estimators (\cite*{wasserman2006all,ruppert2003semiparametric}).


\section{Simulation Study}

Simulation studies were carried out to evaluate the performance of the proposed methods. As a comparison, we also calculated the estimator modeling the survival time and longitudinal surrogate outcome separately based on a Cox model and mixed effects model using the \texttt{R} function \texttt{lme}, denoted as MM in our results below, and the estimator based on the parametric shared parameter models using the \texttt{R} function \texttt{jm} from the package \texttt{JMbayes2} (\cite*{rizopoulos2010jm,rizopoulos2014r}), denoted as JM in our results below. 

We examined 3 settings which were selected in an effort to examine varying model complexities for the model of the longitudinal outcome and its relationship with the terminal event. In all settings, the terminal event was generated from
$$\log(D/10)=-\eta A_i+\epsilon_i,$$
where $\eta=0.5$, the treatment variable $A_i$ was generated from a Bernoulli distribution with success probability 0.5, and $\epsilon_i$ was generated from an extreme value distribution. This resulted in a hazard ratio $\eta=0.5$ and a cumulative hazard function $\Lambda_0(t)=t/10.$ The censoring time was defined as $C_i=C^*_i \wedge \tau$, where $\tau=15$ and $C^*_i$ followed a uniform distribution on (5, 25). The censoring probability for the terminal event $D$ was approximately $25\%$. The observation times were generated from a Poisson process with the intensity function $\mu_{0,i}(t)=20$ for $t=0, 1, 2, ...$, $\mu_{0,i}(t)=0$ for other times. Each individual had 1 observation around each integer time in general and 6 total observations on average. The simulations were constructed such that the longitudinal marker observations for the majority of subjects were terminated by the terminal event $D$ (instead of the censoring time $C$) to highlight the performance of the proposed method in a setting where there was a high probability of informative termination. Results were summarized for sample size $n=200$. 

The settings varied in terms of how the longitudinal variable was generated. We aimed to closely mirror the real clinical trial data. In setting (1), the longitudinal response variable was generated from the model:
\begin{eqnarray*}
(1)\ Y_i(t) = b_{0i} + b_{1i}  (t/4)  + \beta_1 A_i  (t/4)+ e_i(t),
\end{eqnarray*}
where
$b_{0i}$ was the maximum of a normal variable from $N(mean=50, sd=16)$ and 15; $b_{1i}$ was generated from a normal distribution $N(mean=-2, sd=2.75)$, and $e_{i}(t)$ from  a normal distribution $N(mean=0, sd=|0.667 \{b_{0i} + b_{1i}  (t/4) + \beta_1  A_i  (t/4)\}|^{1/2})$. For MM and JM approaches, we included the intercept, t, A and At terms in the mixed effect models (corresponding to the true model) to see the performance. 

In the setting (2), the data were generated from the model:
\begin{eqnarray*}
(2)\ Y_i(t) = b_{0i} + (b_{1i}-5v_i)  (t/4)  + \beta_1 A_i  (t/4)+ e_i(t),
\end{eqnarray*}
where $v_i=\exp{(\epsilon_i)}$ with mean and variance 1, $b_{0i}$ and $b_{1i}$ were generated the same as setting (1), and $e_{i}(t)$ from a normal distribution $N(mean=0, sd=|0.667 \{b_{0i} + (b_{1i}-5v_i) (t/4) + \beta_1  A_i  (t/4)\}|^{1/2})$. Setting (1) represents a scenario in which there was no relationship between the longitudinal outcome and the terminal event, whereas in setting (2) there was relationship through $v_i$. For MM and JM approaches, we also included the intercept, t, A and At terms in the mixed effect models, which is no longer correct due to the specification of the relationship between the longitudinal outcome and the terminal event in the above model.

 In setting (3), the longitudinal outcome was made to be \textit{nonlinear} over time from the model: 
\begin{eqnarray*}
(3)\ Y_i(t)&=&u_i+0.2 v_i t+\beta_0A_i+ A_i g(t)+e_i(t), \text{where}\ g(t)=10\log(1+t).
\end{eqnarray*}
where $u_i$ followed the exponential distribution with scale 1,  $v_i=\exp{(\epsilon_i)}$ with mean and variance 1, $e_i(t)$ was normal with mean $\phi_i$ and standard deviation $0.2t$, and $\phi_i$ was the standard normal random variable.

We examined our proposed SJM approach described in Section \ref{optimal}, the MM approach, and JM approach in settings (1)-(2) when the longitudinal outcome was linear. We examined our proposed SJM approach with nonlinear regression described in Section \ref{expand} in setting (3) when the longitudinal outcome was nonlinear over time. 

\bigskip
\begin{table}[ht]
\centering
\small
\begin{tabular}{rrrrrr|rrr|rrr}
  \hline
(1) &TRUE & EST & $\ESE_{\tiny{\ASE}}$ & CP & MSE& MM &ESE & MSE& JM & ESE   & MSE \\
  \hline
$\eta$ & 0.500 & 0.483 & $0.161_{0.163}$ & 0.956 & 0.026 & 0.482 & 0.163 & 0.027 & 0.500 & 0.277 & 0.076 \\ 
$\beta_0$ & 0.000 & -0.032 & $1.725_{1.780}$ & 0.951 & 2.970 & -0.091 & 1.466 & 2.153 & 0.339 & 1.769 & 3.236 \\ 
$\beta_1/4$ & 2.000 & 2.164 & $0.491_{0.568}$ & 0.977 & 0.268 & 1.996 & 0.273 & 0.074 & 1.916 & 0.313 & 0.105 \\ 
     \hline
(2) &TRUE & EST & $\ESE_{\tiny{\ASE}}$ & CP & MSE& MM &ESE & MSE& JM & ESE   & MSE \\
  \hline
$\eta$ & 0.500 & 0.484 & $0.162_{0.162}$ & 0.944 & 0.026 &  0.483 & 0.162 & 0.026 & 0.029 & 0.250 & 0.285 \\ 
$\beta_0$ & 0.000 & 0.321 & $1.756_{1.807}$ & 0.950 & 3.180 & 0.630 & 1.578 & 2.881 & 1.337 & 1.817 & 5.083 \\ 
$\beta_1/4$ & 2.000 & 2.000 & $0.637_{0.704}$ & 0.960 & 0.406 & {\color{blue}1.300} & 0.427 & 0.673 & {\color{blue}1.260} & 0.453 & 0.752 \\ 
   \hline
\end{tabular}
\caption{Truth and estimates of $\eta, \beta_0, \beta_1$ along with their empirical standard errors (ESE) and mean squared errors (MSE) under settings 1 and 2, the first set of columns shows results for our proposed SJM approach, while the middle and end set of columns show results for the mixed effects model (MM) and the parametric shared parameters approach (JM) approach, respectively; for our proposed approach, we also present the average of the estimated standard errors (ASE) along with the empirical coverage probabilities (CP) of the 95$\%$ confidence intervals.} \label{simu}
\end{table}

Estimation results for Settings (1)-(2) are summarized in Table \ref{simu}. Recall that while we show results for multiple parameters, the primary interest is in estimation of $\beta_1$, the treatment effect on the longitudinal outcome. In setting (1) with correctly specified models in MM and JM, all three methods performed well with small biases, while the MM and JM have smaller empirical standard errors (ESE) and mean squared errors (MSE) compared to the proposed method. Notably, Setting (1) reflects an ideal case that the longitudinal outcome and the survival time are independent, which is not true for our application of interest with GFR and kidney failure or death. Thus, these results are expected since the models are correctly specified and these two estimators are obtained by maximizing the likelihood functions. In Settings (2), the MM estimates of $\beta_1$ are biased with larger MSEs compared to the proposed method. In addition, the JM estimates of $\beta_1$ are all biased with large MSEs. This is expected because the specified models of MM and JM are not correct. Thus, the proposed method is robust to model specification. In these two settings, the proposed estimates have small biases, the average of the estimated standard errors (ASE) are close to the ESEs, the empirical coverage probabilities (CP) are close to the nominal level 95$\%$. 

\begin{figure}[htbp]
\begin{center}
\includegraphics[width=2.25in]{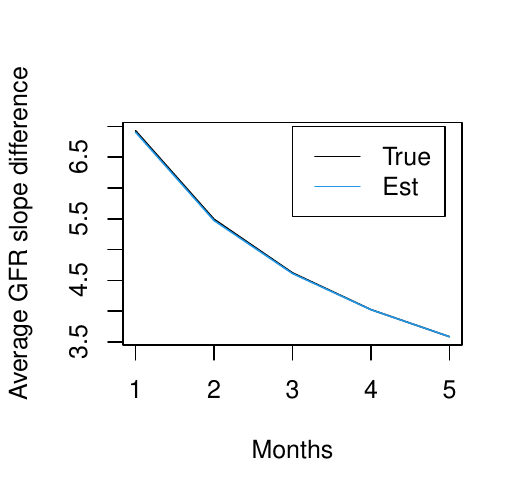}\includegraphics[width=2.25in]{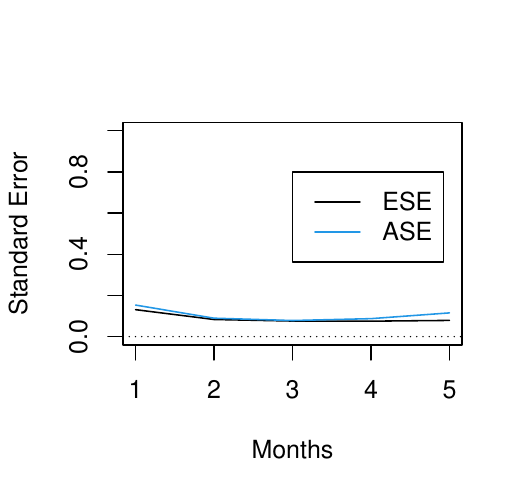}\includegraphics[width=2.25in]{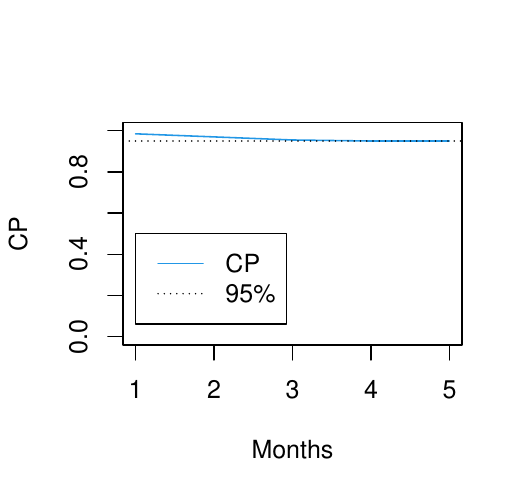}
\caption{Estimation of the treatment effect on the longitudinal outcome slope, $\Delta_{Yslope}(t)$(first panel), average standard error (ASE) and empirical standard error (ESE) estimates (middle panel), and coverage probability (CP) (last panel) over time using our proposed SJM approach with nonlinear regression.}  \label{nonlinear}
\end{center}
\end{figure}

In setting (3), the true value treatment effects: $\Delta_{Y}(t)=E[Y^{(1)}(t)-Y^{(0)}(t)]=\beta_{0}+10\log(1+t)$ and $\Delta_{Yslope}(t)=E[Y^{(1)}(t)-Y^{(1)}(0)-\{Y^{(0)}(t)-Y^{(0)}(0)\}]/t=10\log(1+t)/t $, are not constant over time. Of course, the proposed estimation methods under linear model, MM, and JM would all produce one constant estimate of the treatment effect for all time t, by design. Therefore, we only report results for our proposed approach from using nonlinear regression to estimate $\Delta_{Yslope}(t)$, described Section \ref{expand}. Results are shown in Figure \ref{nonlinear}, which show that the estimated treatment effects on the longitudinal outcome slope over time are close to the true values with small biases, the ASEs are close to corresponding ESEs, and the CPs are close to the nominal level 95$\%$.


\section{Real Data Analysis} \label{application}

The Reduction of Endpoints in NIDDM with the Angiotensin II Antagonist Losartan (RENAAL) Study was an investigator-initiated, multinational, double-blind, randomized, placebo-controlled study designed to evaluate the renoprotective effects of losartan compared with placebo in 1513 patients with type 2 diabetes and nephropathy (\cite*{brenner2002effects}). GFR were measured at visit time points, which were scheduled every three months with a mean follow-up time of 3.4 years (range, 2.3 to 4.6). The terminal event was death or kidney failure. 
 
We first applied our proposed SJM linear method from Section \ref{optimal} to jointly model the longitudinal outcome GFR and the terminal event under model (\ref{gfr}).  The estimate of the treatment effect on the GFR slope ($\widehat{\beta}_1=0.110$ with estimated standard error 0.038), our primary quantity of interest, is positive and significant, meaning that the GFR trajectory for the treatment group decreased less steeply, or the GFR in the control group decreased more steeply. This result is consistent with MM and JM estimates, and also previous findings from the literature (\cite*{inker2023meta}).

\begin{table}[ht]
\centering
\begin{tabular}{rrrrrrr}
  \hline
  & EST & eSE & MM & eSE  & JM & eSE \\
  \hline
$\eta$ & -0.198 & 0.083 & -0.198 & 0.086& -0.093 & 0.113 \\ 
$\beta_0$ & -0.221 & 0.842 & -0.986 & 0.710 & -0.968 & 0.711\\ 
$\beta_1$ & 0.110 & 0.038 & 0.059 & 0.023& 0.058 & 0.024 \\ 
   \hline
\end{tabular}
\caption{Estimates of $\eta, \beta_0, \beta_1$ along with the estimated standard errors (eSE) applying the proposed SJM linear approach to the RENAAL Study.}
\label{real}
\end{table}

\begin{figure}[htbp]
\begin{center}
\includegraphics[width=3.5in]{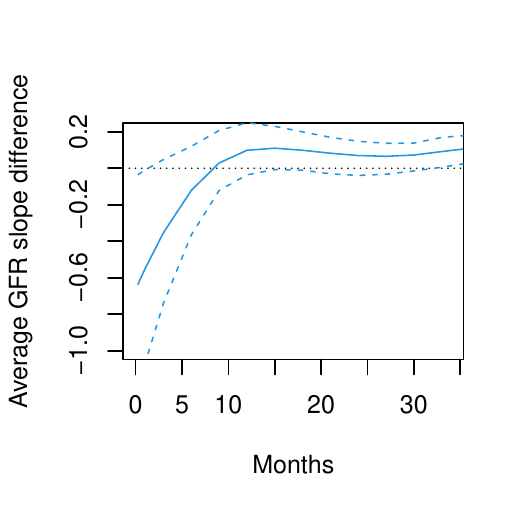}
\caption{Estimates (solid line) and 95\% pointwise confidence bands (dashed lines) of the treatment effect on GFR slope over time applying the proposed SJM approach with nonlinear regression to the RENAAL Study.}   \label{sloppdiff}
\end{center}
\end{figure}

 Of course, we suspect that model (\ref{gfr}) is likely untrue for GFR; thus, we next applied our proposed SJM method with nonlinear regression from Section \ref{expand}. Figure \ref{sloppdiff} shows these results, demonstrating that the estimated treatment effect on the longitudinal outcome slope is negative in the beginning, and then positive later in time, implying that there is a steeper decline in GFR for the treatment group early in time but overall, the decline in GFR for the treatment group is slower and smaller in magnitude compared to the control group. Thus, this approach offers a unique and robust way to estimate a nonlinear treatment effect, which is particularly useful in CKD trials examining GFR.

\section{Discussion}
Motivated by the use of GFR as a surrogate outcome in CKD clinical trials, we have built a semiparametric framework to jointly model a longitudinal outcome and a terminal event in order to estimate the treatment effect on the longitudinal outcome trajectory. Our SJM method offers several unique contributions compared to current available methods. First, the proposed model is much more flexible compared to the widely used shared parameter models, where the model for the longitudinal outcome, the model for the terminal event, and the relationship between the longitudinal outcome and the terminal event are all parametric. In contrast, our proposed models for the longitudinal outcome and terminal events are semiparametric, and the relationship between the longitudinal outcome and the terminal event is nonparametric. Second, the proposed SJM can be extended to include accommodate a nonlinear trajectory of the longitudinal outcome in a straightforward way, while such an extension would be difficult under the existing parametric framework. Third, while the estimation under the traditional parametric framework is usually computationally intensive and may not converge, we instead proposed an estimating equation based method to estimate the treatment effect on the slope of the longitudinal outcome, which has a nice closed form. Lastly, the proposed model and estimation are distinct from existing semiparametric methods for a longitudinal outcome by modeling the longitudinal surrogate while adjusting for informative termination by the terminal event. 

To the best of our knowledge, this is the first work to use a flexible and robust semiparametric method to analyze the longitudinal surrogate marker, GFR slope, in CKD clinical trials. Notably, this statistical methodology is important not only for analysis of GFR slope, but also builds a framework that is directly translatable to other treatments, biomarkers, and conditions. 

While our proposed method can certainly be used to estimate the treatment effect on a general longitudinal outcome, care is needed when making any decisions about the treatment based on this estimate if the marker is not a good surrogate. If the goal is to make inference about the treatment effect on the primary outcome based on the treatment effect on the longitudinal marker, then the longitudinal marker must be a validated surrogate. Certainly, many methods exist to validate a surrogate. A recent review by \cite{elliott2023surrogate} describes existing frameworks for surrogacy validation including the proportion of treatment effect explained framework, the principal stratification framework, and the meta-analytic framework. For a longitudinal marker within a singe clinical trial in particular, \cite{agniel2021evaluation} have proposed multiple approaches to estimate the proportion of the treatment effect explained by a longitudinal marker, one based on a functional linear model, another on a generalized additive model, and a kernel-based approach with implementation via the \texttt{longsurr R} package on CRAN. Our particular interest in estimating the effect of a treatment on GFR is to potentially pool such estimates across multiple CKD trials via a meta-analysis. Currently, CKD trials tend to report treatment effects on GFR that are estimated via parametric approaches. Of course, when these parametric assumptions do not hold for some or all of the trials, it is not reasonable to pool such estimates. Our proposed approach offers a more flexible estimation method to enable future meta-analyses that require future assumptions.

Our proposed approach has some limitations. Under the semiparametric framework, the efficiency of the estimator for the treatment effect on the longitudinal outcome is limited which is directly due to the semiparametric nature and estimating equation based method (\cite*{lin2001semiparametric}). Although in general a nonparametric method is less efficient compared to a parametric method, it can provide better estimates with large samples and unknown distributions. Nonparametric methods are preferred when flexibility, robustness, and fewer assumptions about the underlying data distribution are desired, especially in the presence of complex data structures or when the true distribution is unknown. To obtain a more efficient estimator of $\bbeta$, one could consider a weighted estimating equation by adding weight to the estimating equation (\ref{estimating}), with some prior knowledge of the covariance of the longitudinal outcome. This warrants future research. The proposed approach, while not intended to be sufficiently efficient relative to parametric methods, is a step towards development of methods that balance the robustness achieved by more flexible approaches and the efficiency of parametric models.





\bigskip

{\noindent {\bf\large Acknowledgements}

We thank Dr. Hiddo Lambers Heerspink as representative of the RENAAL investigators for approving the use of the RENAAL trial data. This work was supported by the National Institute of Diabetes and Digestive and Kidney Diseases (R01DK118354) and the National Natural Science Foundation of China (12171329).  
 
\bibliographystyle{biometri}
\bibliography{ref}

\begin{thebibliography}{}

\bibitem[\protect\astroncite{Agniel \&\ Parast}{Agniel \&\
  Parast}{2021}]{agniel2021evaluation}
{\sc Agniel, D. \&\ Parast, L.} (2021).
\newblock Evaluation of longitudinal surrogate markers.
\newblock {\em Biometrics} {\bf 77}, 477--489.

\bibitem[\protect\astroncite{Brenner, Cooper, de~Zeeuw, Investigators
  et~al.}{Brenner et~al.}{2002}]{brenner2002effects}
{\sc Brenner, B., Cooper, M., de~Zeeuw, D., Investigators, R.~S. et~al.}
  (2002).
\newblock Effects of losartan on renal and cardiovascular outcomes in patients
  with type 2 diabetes and nephropathy.
\newblock {\em ACC Current Journal Review} {\bf 1}, 26.

\bibitem[\protect\astroncite{Elliott}{Elliott}{2023}]{elliott2023surrogate}
{\sc Elliott, M.~R.} (2023).
\newblock Surrogate endpoints in clinical trials.
\newblock {\em Annual Review of Statistics and its Application} {\bf 10},
  75--96.

\bibitem[\protect\astroncite{FDA}{FDA}{2024}]{FDA}
{\sc FDA} (2024).
\newblock Table of surrogate endpoints that were the basis of drug approval or
  licensure.
\newblock {
  https://www.fda.gov/drugs/development-resources/table-surrogate-endpoints-were-basis-drug-approval-or-licensure}.

\bibitem[\protect\astroncite{Fleming \&\ Harrington}{Fleming \&\
  Harrington}{2013}]{fleming}
{\sc Fleming, T.~R. \&\ Harrington, D.~P.} (2013).
\newblock {\em Counting processes and survival analysis}, volume 625.
\newblock John Wiley \& Sons.

\bibitem[\protect\astroncite{Inker, Collier, Greene, Miao, Chaudhari, Appel,
  Badve, Caravaca-Font{\'a}n, Del~Vecchio, Floege et~al.}{Inker
  et~al.}{2023}]{inker2023meta}
{\sc Inker, L.~A., Collier, W., Greene, T., Miao, S., Chaudhari, J., Appel,
  G.~B., Badve, S.~V., Caravaca-Font{\'a}n, F., Del~Vecchio, L., Floege, J.
  et~al.} (2023).
\newblock A meta-analysis of gfr slope as a surrogate endpoint for kidney
  failure.
\newblock {\em Nature Medicine} pages 1--10.

\bibitem[\protect\astroncite{Jin, Liu, Albert \&\ Ying}{Jin
  et~al.}{2006}]{jin2006analysis}
{\sc Jin, Z., Liu, M., Albert, S. \&\ Ying, Z.} (2006).
\newblock Analysis of longitudinal health-related quality of life data with
  terminal events.
\newblock {\em Lifetime Data Analysis} {\bf 12}, 169--190.

\bibitem[\protect\astroncite{Li, Astor, Lewis, Hu, Appel, Lipkowitz, Toto,
  Wang, Wright~Jr \&\ Greene}{Li et~al.}{2012}]{li2012longitudinal}
{\sc Li, L., Astor, B.~C., Lewis, J., Hu, B., Appel, L.~J., Lipkowitz, M.~S.,
  Toto, R.~D., Wang, X., Wright~Jr, J.~T. \&\ Greene, T.~H.} (2012).
\newblock Longitudinal progression trajectory of gfr among patients with ckd.
\newblock {\em American journal of kidney diseases} {\bf 59}, 504--512.

\bibitem[\protect\astroncite{Lin \&\ Ying}{Lin \&\
  Ying}{2001}]{lin2001semiparametric}
{\sc Lin, D.~Y. \&\ Ying, Z.} (2001).
\newblock Semiparametric and nonparametric regression analysis of longitudinal
  data.
\newblock {\em Journal of the American Statistical Association} {\bf 96},
  103--126.

\bibitem[\protect\astroncite{Parast, McDermott \&\ Tian}{Parast
  et~al.}{2016}]{parast2016robust}
{\sc Parast, L., McDermott, M.~M. \&\ Tian, L.} (2016).
\newblock Robust estimation of the proportion of treatment effect explained by
  surrogate marker information.
\newblock {\em Statistics in medicine} {\bf 35}, 1637--1653.

\bibitem[\protect\astroncite{Rizopoulos}{Rizopoulos}{2010}]{rizopoulos2010jm}
{\sc Rizopoulos, D.} (2010).
\newblock Jm: An r package for the joint modelling of longitudinal and
  time-to-event data.
\newblock {\em Journal of statistical software} {\bf 35}, 1--33.

\bibitem[\protect\astroncite{Rizopoulos}{Rizopoulos}{2014}]{rizopoulos2014r}
{\sc Rizopoulos, D.} (2014).
\newblock The r package jmbayes for fitting joint models for longitudinal and
  time-to-event data using mcmc.
\newblock {\em arXiv preprint arXiv:1404.7625} .

\bibitem[\protect\astroncite{Ruppert et~al.}{Ruppert
  et~al.}{2003}]{ruppert2003semiparametric}
{\sc Ruppert, D., Wand, M.~P. \&\ Carroll, R.~J.} (2003).
\newblock {\em Semiparametric regression}.
\newblock Number~12. Cambridge university press.

\bibitem[\protect\astroncite{Sun, Park, Sun \&\ Zhao}{Sun
  et~al.}{2005}]{sun2005semiparametric}
{\sc Sun, J., Park, D.-H., Sun, L. \&\ Zhao, X.} (2005).
\newblock Semiparametric regression analysis of longitudinal data with
  informative observation times.
\newblock {\em Journal of the American Statistical Association} {\bf 100},
  882--889.

\bibitem[\protect\astroncite{Sun, Sun \&\ Liu}{Sun
  et~al.}{2007}]{sun2007regression}
{\sc Sun, J., Sun, L. \&\ Liu, D.} (2007).
\newblock Regression analysis of longitudinal data in the presence of
  informative observation and censoring times.
\newblock {\em Journal of the American Statistical Association} {\bf 102},
  1397--1406.

\bibitem[\protect\astroncite{Sun, Song, Zhou \&\ Liu}{Sun
  et~al.}{2012}]{sun2012joint}
{\sc Sun, L., Song, X., Zhou, J. \&\ Liu, L.} (2012).
\newblock Joint analysis of longitudinal data with informative observation
  times and a dependent terminal event.
\newblock {\em Journal of the American Statistical Association} {\bf 107},
  688--700.

\bibitem[\protect\astroncite{van~der Vaart \&\ Wellner}{van~der Vaart \&\
  Wellner}{1996}]{VW}
{\sc van~der Vaart, A. \&\ Wellner, J.} (1996).
\newblock {\em Weak convergence and empirical processe}.
\newblock New York: Springer.

\bibitem[\protect\astroncite{Vonesh, Tighiouart, Ying, Heerspink, Lewis,
  Staplin, Inker \&\ Greene}{Vonesh et~al.}{2019}]{vonesh2019mixed}
{\sc Vonesh, E., Tighiouart, H., Ying, J., Heerspink, H.~L., Lewis, J.,
  Staplin, N., Inker, L. \&\ Greene, T.} (2019).
\newblock Mixed-effects models for slope-based endpoints in clinical trials of
  chronic kidney disease.
\newblock {\em Statistics in medicine} {\bf 38}, 4218--4239.

\bibitem[\protect\astroncite{Vonesh, Greene \&\ Schluchter}{Vonesh
  et~al.}{2006}]{vonesh2006shared}
{\sc Vonesh, E.~F., Greene, T. \&\ Schluchter, M.~D.} (2006).
\newblock Shared parameter models for the joint analysis of longitudinal data
  and event times.
\newblock {\em Statistics in medicine} {\bf 25}, 143--163.

\bibitem[\protect\astroncite{Wang, Parast, Tian \&\ Cai}{Wang
  et~al.}{2020}]{wang2020model}
{\sc Wang, X., Parast, L., Tian, L. \&\ Cai, T.} (2020).
\newblock Model-free approach to quantifying the proportion of treatment effect
  explained by a surrogate marker.
\newblock {\em Biometrika} {\bf 107}, 107--122.

\bibitem[\protect\astroncite{Wasserman}{Wasserman}{2006}]{wasserman2006all}
{\sc Wasserman, L.} (2006).
\newblock {\em All of nonparametric statistics}.
\newblock Springer Science \& Business Media.

\bibitem[\protect\astroncite{Xie, Bowe, Xian, Balasubramanian \&\ Al-Aly}{Xie
  et~al.}{2016}]{xie2016estimated}
{\sc Xie, Y., Bowe, B., Xian, H., Balasubramanian, S. \&\ Al-Aly, Z.} (2016).
\newblock Estimated gfr trajectories of people entering ckd stage 4 and
  subsequent kidney disease outcomes and mortality.
\newblock {\em American Journal of Kidney Diseases} {\bf 68}, 219--228.

\end{thebibliography}

\clearpage
\section*{Appendix} \label{app1}
\vspace{0.1in}

In order to study the asymptotic distribution of  $\widehat{\bbeta}$, we require the following regularity conditions:
\begin{enumerate}
\item[](C1) \ $\{Y_i(\cdot), N_i(\cdot), T_i, \delta_i, \bZ_i\}, i=1,..., n $,
 are independent and identically distributed.

\item[](C2) \ $N(\tau)$ and $\Z$ are bounded almost surely, \
$Y(t)$ is of bounded variation and \ $P(T \ge \tau)>0$.

\item[](C3) \  $A$ is nonsingular, where $
  A= E\Big[\int_0^{\tau}\{\tilde{\bZ}_i(t)-\bar{\bz}(t, \bZ_i)\}^{\otimes 2}
I(T_i\ge t)d \mu_0(t)\Big]
$  and $\bar{\bz}(t, \bZ_i)$ is the limit of $\bar\bZ_i(t)$.
\end{enumerate}

\vskip 0.5cm
\noindent\textbf{Proof of (\ref{dH})}\\
First we show that
\begin{eqnarray*}
&&E[ \{Y(t)-{\bbeta}^{\trans}\tilde{\bZ}(t)\}dN(t)|\bZ,T>t]\\
&&=E[ \{Y(t)-{\bbeta}^{\trans}\tilde{\bZ}(t)\}dN(t)|\bZ,D>t]\\
&&=E\left(E\left[ \{Y(t)-{\bbeta}^{\trans}\tilde{\bZ}(t)\}dN(t)|\bZ,v,D>t\right]|\bZ,D>t\right)\\
&&=E\left[  \alpha_0(t,v)d\mu_0(t) |\bZ,D>t\right]\\
&&= {E\left[ d\Asc(t,v) |\bZ,\log \Lambda_0(D)+\Beta^{\trans}\bZ>=\log \Lambda_0(t)+\Beta^{\trans}\bZ \right]}\\
&&= {E\left[ d\Asc(t,v) |\varepsilon>s \right]|_{s=\log \Lambda_0(t)+\Beta^{\trans}\bZ}}\\
&&= dH(t,\log \Lambda_0(t)+\Beta^{\trans}\bZ).
\end{eqnarray*}
thus we have
\begin{eqnarray*}
 dH(t,s)&&= E\left[ d\Asc(t,v) |\varepsilon>s  \right]\\
&&=E\left[ d\Asc(t,v) |\log \Lambda_0(D)+\Beta^{\trans}\bZ>s  \right]\\
&&=  \frac{E[d\Asc(t,v) I(\log \Lambda_0(D)+\Beta^{\trans}\bZ>s)]}{E[ I(\log \Lambda_0(D)+\Beta^{\trans}\bZ>s)]}\\
&&=  \frac{E[d\Asc(t,v) I(\log \Lambda_0(D)+\Beta^{\trans}\bZ>s>\log \Lambda_0(t)+\Beta^{\trans}\bZ)]}{E[ I(\log \Lambda_0(D)+\Beta^{\trans}\bZ>s>\log \Lambda_0(t)+\Beta^{\trans}\bZ)]}\\
&&=  \frac{E[\{Y(t)-{\bbeta}^{\trans}\tilde{\bZ}(t)\}dN(t) I(\log \Lambda_0(D)+\Beta^{\trans}\bZ>s>\log \Lambda_0(t)+\Beta^{\trans}\bZ)]}{E[ I(\log \Lambda_0(D)+\Beta^{\trans}\bZ>s>\log \Lambda_0(t)+\Beta^{\trans}\bZ)]}\\
\end{eqnarray*}

\noindent \textbf{Prof of  Theorem 1}

\noindent Define
 $$
  d\bar{M}(t, \bZ; \Beta, \Lambda)
 =\frac{\sum_{j=1}^n {\Phi}_j(t, \bZ; \Beta, \Lambda)dM_j(t)}
 {\sum_{j=1}^n \Phi_j(t, \bZ; \Beta, \Lambda)}, \hskip 0cm
 $$
 $$
 d\bar M_0(t, \bZ; \Beta, \Lambda)=\frac{E[{\Phi}_j(t, \bZ; \Beta, \Lambda)dM_j(t)|\bZ]}
 {E[ \Phi_j(t, \bZ; \Beta, \Lambda)|\bZ]}. \hskip 0cm
  $$
  and
 $$
 \bar{\bz}(t, \bZ; \Beta, \Lambda)
 =\frac{E[\tilde{\bZ}_j (t){\Phi}_j(t, \bZ; \Beta, \Lambda)|\bZ]}
 {E[ \Phi_j(t, \bZ; \Beta, \Lambda)|\bZ]},
 $$
 where $dM_i(t)=\{Y_i(t)-\bbeta_0'\tilde{\bZ}_i(t)\}d N_i(t)$ and
 $$
 {\Phi}_i(t, \bZ; \Beta, \Lambda)=I\{\log \Lambda(T_i) +\Beta'\bZ_i \ge \log \Lambda(t)
 +\Beta'\bZ  \ge \log \Lambda(t)+\Beta'\bZ_i\}.
 $$
 Denote $\bar{\bz}(t, \bZ)\equiv \bar{\bz}(t, \bZ;  \Beta_0, \Lambda_0),$
 and ${\Phi}_i(t, \bZ) \equiv {\Phi}_i(t, \bZ; \Beta_0, \Lambda_0).$
 Using the functional delta method (\cite*{VW}, Theorem 3.9.4,
 page 374), we obtain
 $$
n^{-1/2}\sum_{i=1}^n
 \int_0^{\tau}  \big\{\tilde{\bZ}_i(t)-\bar{\bz}(t, \bZ_i)\big\}
I(T_i\ge t) \{d \bar{M}(t, \bZ_i; \widehat \Beta, \widehat\Lambda_0)
 -d \bar{M}_0(t, \bZ_i; \widehat \Beta, \widehat\Lambda_0) \big\}
  $$
$$
 =n^{-1/2}\sum_{i=1}^n
 \int_0^{\tau}  \Big[ \int \big\{\tilde{\bz}(t)-\bar{\bz}(t, \bz)\big\}I(u \ge t)
 \frac{ \Phi_i(t, \bz)}{E\Phi_i(t, \bz)} dF(\bz, u )\Big]
 d M_i(t)
  $$
 $$
 -n^{-1/2}\sum_{i=1}^n \int \Big[
 \int_0^{\tau}  \big\{\tilde{\bz}(t)-\bar{\bz}(t, \bz)\big\}I(u \ge t)\Phi_i(t, \bz)
 \hskip 2.5cm
 $$
 $$
 \times \frac{E[\Phi_i(t, z) d M_i(t)] }{(E[\Phi_i(t, z)])^2} \Big]
 dF(\bz, u) +o_p(1),  \hskip 3.2cm
 \eqno (A.1)
  $$
  where  ${F}(\bz, u)$ is  the joint probability
measure of $(\bZ_i, T_i).$
Note that by \cite{fleming},
$$ \widehat{\Beta}-\Beta_0= \Omega^{-1}n^{-1}\sum_{i=1}^n \int_0^{\tau}
 \big\{\bZ_i-\bar{\bz}^D(t) \big\}dM_i^{D} (t)+o_p(n^{-1/2}),
 $$
 and
 $$
 \widehat{\Lambda}_0(t)-\Lambda_0(t) = n^{-1}\sum_{i=1}^n \int_0^{t}
\frac{dM_i^{D}(u)}{{s}^{(0)}(u; \Beta_0)}
 -\int_0^{t}\bar{\bz}^D(u)'d\Lambda_0(u)
 (\widehat{\Beta}-\Beta_0)+o_p(n^{-1/2}),
$$
where
$ M_i^{D}(t)=I(T_i\ge t,\delta_i=1)-\int_0^t
I(T_i\ge u)\exp\{\Beta_0'\bZ_i\}d\Lambda_0(u),
$
 and $\Omega,$ ${s}^{(0)}(t;  \Beta_0)$ and $\bar{\bz}^D(t)$ are the limits of
 $\widehat{\Omega},$ ${S}^{(0)}(t; \Beta_0)$
 and $\bar{\bZ}^D(t; \Beta_0),$ respectively.
 Let $dR_{\Beta}(t, \bZ)$ and $dR_{\Lambda}(t, \bZ)$ be the derivative
and the Hadamard derivative of $d\bar{M}_0(t, \bZ; \Beta_0, {\Lambda}_0)$
with respect to $\Beta$ and $\Lambda$, respectively.  Then we have
 $$
 n^{-1/2}\sum_{i=1}^n
 \int_0^{\tau}  \big\{\tilde{\bZ}_i(t)-\bar{\bz}(t, \bZ_i)\big\}
I(T_i\ge t)  \big\{d\bar{M}_0(t, \bZ_i; \widehat\Beta, \widehat{\Lambda}_0)
 -d\bar{M}_0(t, \bZ_i; \Beta_0, {\Lambda}_0) \big\}
 $$
$$
= n^{-1/2}\sum_{i=1}^n \int_0^{\tau}
 \Big[P_2 \Omega^{-1} \big\{\bZ_i-\bar{\bz}^D(t) \big\}+\frac{H_2(t)}{{s}^{(0)}(t; \Beta_0)}
 \Big]dM_i^{D} (t)+o_p(1), \hskip 2.3cm
 \eqno (A.2)
 $$
where $$
 H_2(t)=E\Big[ \int_t^{\tau}  \big\{\tilde{\bZ}_i(u)-\bar{\bz}(u, \bZ_i)\big\}I(T_i\ge u)dR_{\Lambda}(u, \bZ_i)\Big],
$$
 $$
{P}_2=E\Big[\int_0^{\tau}  \big\{\tilde{\bZ}_i(t)-\bar{\bz}(t,
\bZ_i)\big\}I(T_i\ge t) \Big\{ dR_{\Beta}(t, \bZ_i)
\hskip 0.8cm $$
$$
-\Big(\int_0^{t}\bar{\bz}^D(u)'d\Lambda_0(u)\Big) dR_{\Lambda}(t, \bZ_i)\Big\}\Big].
\hskip 1cm $$
In addition, it is easy to derive that
 $$
 n^{-1/2} U(\bbeta_0)=n^{-1/2} \sum_{i=1}^n
 \int_0^{\tau} \big\{\tilde{\bZ}_i(t)-\bar{\bZ}_i(t)\big\}
I(T_i\ge t) \Big[ d M_i(t) -d\bar{M}_0(t, \bZ_i; \Beta_0, \Lambda_0) $$
  $$
\hskip 3cm -\big\{d \bar{M}(t, \bZ_i; \widehat\Beta, \widehat{\Lambda}_0)
 - d\bar{M}_0(t, \bZ_i; \widehat\Beta, \widehat{\Lambda}_0) \big\} \hskip 2.3cm
   $$
  $$
 \hskip 2.6cm  - \big\{d \bar{M}_0(t, \bZ_i; \widehat\Beta, \widehat{\Lambda}_0)
  - d\bar{M}_0(t, \bZ_i; \Beta_0, \Lambda_0) \big\} \Big]+o_p(1) \eqno (A.3)
  $$
  Since $\sup_{i, t}|\bar{\bZ}_i(t)-\bar{\bz}(t, \bZ_i)|
 \rightarrow 0,$
 it follows from (A.1), (A.2) and (A.3) that
 $$
  n^{-1/2} U(\bbeta_0)=n^{-1/2}\sum_{i=1}^n \bxi_i+o_p(1), \hskip 5.6cm \eqno (A.4)
 $$
 where
 $$
 \bxi_i=\int_0^{\tau}  \big\{\tilde{\bZ}_i(t)-\bar{\bz}(t, \bZ_i)\big\}
 I(T_i\ge t) \Big[d M_i(t)-d\bar{M}_0(t, \bZ_i; \Beta_0, \Lambda_0)\Big] \hskip 1.8cm
 $$
 $$
 - \int_0^{\tau}  \Big[ \int \big\{\tilde{\bz}(t)-\bar{\bz}(t, \bz)\big\}I(u \ge t)
 \frac{ \Phi_i(t, \bz)}{E[\Phi_i(t, \bz)]}   dF(\bz, u)\Big]d M_i(t)
  \hskip 0cm $$
 $$
  + \int \Big[
 \int_0^{\tau}  \big\{\tilde{\bz}(t)-\bar{\bz}(t, z)\big\}I(u \ge t)\Phi_i(t, \bz)
 \hskip 3.5cm $$
 $$
\times  \frac{E[\Phi_i(t, \bz) d M_i(t)] }{(E\Phi_i(t, \bz))^2} \Big] dF(\bz, u)
 \hskip 5cm $$
 $$
 - \int_0^{\tau}
 \Big[P_2 \Omega^{-1} \big\{\bZ_i-\bar{\bz}^D(t) \big\}+\frac{H_2(t)}{{s}^{(0)}(t; \Beta_0)}
 \Big]dM_i^{D} (t). \hskip 4.2cm
 $$
Note that
$-n^{-1}\partial {U}(\bbeta_0)/\partial \bbeta$
converge in probability to $A$.
It then follows from (A.4) and the Taylor series expansion
that $$
n^{1/2} (\widehat \bbeta-\bbeta_0) =A^{-1}n^{-1/2}  {U}(\bbeta_0)+o_p(1)=A^{-1}n^{-1/2}\sum_{i=1}^n \bxi_i  +o_p(1),
$$
 which means that $n^{1/2} ( \widehat \bbeta - \bbeta_0 )$
  have asymptotically a  normal distribution with mean zero and
  covariance matrix $A^{-1}\Sigma (A^{\trans})^{-1}$,
   where $\Sigma=E[\bxi_i^{\otimes 2}].$

\end{document}